\documentclass{svjour2}
\usepackage{graphicx}

\journalname{General Relativity and Gravitation}

\def\beq{\begin{equation}}
\def\eeq{\end{equation}}

\def\cstok#1{\leavevmode\thinspace\hbox{\vrule\vtop{\vbox{\hrule\kern1pt
\hbox{\vphantom{\tt/}\thinspace{\tt#1}\thinspace}}
\kern1pt\hrule}\vrule}\thinspace}

\begin{document}

\title{Solution of Maxwell's equations on a de Sitter background}

\author{Donato Bini \and
        Giampiero Esposito \and
        Roberto Valentino Montaquila  
}

\institute{Donato Bini 
              \at
Istituto per le Applicazioni del Calcolo ``M. Picone,'' CNR, 
Via del Policlinico 137, I-00161 Rome, Italy\\
ICRA, University of Rome ``La Sapienza,'' I--00185 Rome, Italy\\
Istituto Nazionale di Fisica Nucleare, 
Sezione di Firenze, Polo Scientifico, Via Sansone 1, I--50019, 
Sesto Fiorentino (FI), Italy\\
              \email{binid@icra.it} 
    \and
              Giampiero Esposito 
              \at
Istituto Nazionale di Fisica Nucleare, Sezione di
Napoli, Complesso Universitario di Monte S. Angelo, Via Cintia,
Edificio 6, 80126 Napoli, Italy\\
                            \email{giampiero.esposito@na.infn.it}         
    \and
          Roberto Valentino Montaquila 
              \at
Dipartimento di Scienze Fisiche, Universit\`a di Napoli
Federico II, Complesso Universitario di Monte S. Angelo,
Via Cintia, Edificio 6, 80126 Napoli, Italy\\
Istituto Nazionale di Fisica Nucleare, Sezione di
Napoli, Complesso Universitario di Monte S. Angelo, Via Cintia,
Edificio 6, 80126 Napoli, Italy
              \email{montaquila@na.infn.it}
}

\date{Received: date / Accepted: date / Version: \today}

\maketitle

\begin{abstract}
The Maxwell equations for the electromagnetic potential, 
supplemented by the Lorenz gauge condition,
are decoupled and solved exactly in de Sitter space-time studied in static
spherical coordinates. There is no source besides the background. 
One component of the vector field is expressed,
in its radial part, through the solution of a fourth-order 
ordinary differential equation obeying given initial conditions. 
The other components of the vector field are then found by acting
with lower-order differential operators on the solution of the 
fourth-order equation (while the transverse part is
decoupled and solved exactly from the beginning). The whole four-vector
potential is eventually expressed through hypergeometric functions and
spherical harmonics. Its radial part is plotted for given choices of
initial conditions. We have thus completely succeeded in solving the
homogeneous vector wave equation for Maxwell theory in the Lorenz gauge
when a de Sitter spacetime is considered, which is relevant both for
inflationary cosmology and gravitational wave theory. The decoupling
technique and analytic formulae and plots are completely original. 
This is an important step
towards solving exactly the tensor wave equation in de Sitter 
space-time, which has important applications to the theory of gravitational
waves about curved backgrounds.  
\end{abstract}

\section{Introduction}

It is by now well known that the problem of solving vector and tensor 
wave equations in curved spacetime, motivated by physical problems 
such as those occurring in gravitational wave theory and relativistic
astrophysics, is in general a challenge even for the modern computational
resources. Within this framework, 
a striking problem is the coupled nature of the set of 
hyperbolic equations one arrives at. For example, on using the
Maxwell action functional
\begin{equation}
S=-{1\over 4}\int_{M}F_{ab}F^{ab}\sqrt{-g}\; d^{4}x
\label{(1.1)}
\end{equation}
jointly with the Lorenz \cite{PHMAA-34-287} gauge condition
\begin{equation}
\nabla^{b}A_{b}=0,
\label{(1.2)}
\end{equation}
one gets, in vacuum, the coupled equations for the electromagnetic
potential
\begin{equation}
\left(-\delta_{a}^{\; b}\cstok{\ }+R_{a}^{\; b}\right)A_{b}=0.
\label{(1.3)}
\end{equation}
It was necessary to wait until the mid-seventies to obtain a major
breakthrough in the solution of coupled hyperbolic equations such as 
(3), thanks to the work of Cohen and Kegeles \cite{PHRVA-D10-1070},
who reduced the problem to the task of finding solutions of a complex
scalar equation. Even on considering specific backgrounds such as de
Sitter spacetime, only the Green functions of the wave operator have
been obtained explicitly so far \cite{CMPHA-103-669}, to the best
of our knowledge.

Thus, in a recent paper \cite{00436}, 
we have studied the vector and tensor wave
equations in de Sitter space-time with static spherical coordinates, so
that the line element reads as
\begin{equation}
ds^{2}=-f \; dt^{2} +f^{-1} \; dr^{2}
+r^{2}(d\theta^{2}+\sin^{2}\theta d\phi^{2}), 
\label{(1.4)}
\end{equation}
where $f \equiv 1-H^{2}r^{2}$, and $H$ is the Hubble constant related to
the cosmological constant $\Lambda$ by $H^{2}={\Lambda \over 3}$. The vector
field $X$ solving the vector wave equation can be expanded in spherical
harmonics according to \cite{PHRVA}
\begin{eqnarray}
X&=& {\widetilde Y}_{lm}(\theta)e^{-i(\omega t-m \phi)}
\Bigr[f_{0}(r)dt+f_{1}(r)dr \Bigr] \nonumber \\
&+& e^{-i(\omega t-m \phi)}\left[-{mr \over \sin \theta}f_{2}(r)
{\widetilde Y}_{lm}(\theta)+f_{3}(r){d{\widetilde Y}_{lm}\over d\theta}
\right]d\theta \nonumber \\
&+& i e^{-i(\omega t-m \phi)}\left[-r \sin \theta f_{2}(r)
{d{\widetilde Y}_{lm}\over d\theta}+m f_{3}(r)
{\widetilde Y}_{lm}(\theta)\right]d\phi,
\label{(1.5)}
\end{eqnarray}
where ${\widetilde Y}_{lm}(\theta)$ is the $\theta$-dependent part of the
spherical harmonics $Y_{lm}(\theta,\phi)$. As we have shown in 
\cite{00436}, the function $f_{2}$ is decoupled and obeys a differential
equation solved by a combination of hypergeometric functions, i.e.
\begin{equation}
f_{2}(r)= f^{-i \Omega /2}\biggr[U_{1}r^{l}
F\left(a_{-},a_{+};{3\over 2}+l;H^{2}r^{2}\right)
+ U_{2}r^{-l-1}F\left(a_{+},a_{-};{1\over 2}-l;
H^{2}r^{2}\right)\biggr],
\label{(1.6)}
\end{equation}
where
\begin{equation}
\Omega \equiv {\omega \over H}, \;
a_{\pm} \equiv -{1\over 4}\left({2i \Omega}-3-2l \pm 1 \right).
\label{(1.7)}
\end{equation}

At this stage, however, the problem remained of solving explicitly also
for $f_{0}(r),f_{1}(r),f_{3}(r)$ in the expansion (5). For this
purpose, Sec. II derives the decoupling procedure for such modes in
de Sitter, and Sec. III writes explicitly the decoupled equations.
Section IV solves explicitly for $f_{0},f_{1},f_{3}$ in terms of
hypergeometric functions, while Sec. V plots such solutions for 
suitable initial conditions. Relevant details are presented in the
Appendices. 
It now remains to be seen whether a technique
similar to Secs. II and III can be used to solve completely also the
tensor wave equation obtained in \cite{00436}. 
This technical step would have far
reaching consequences for the theory of gravitational waves in cosmological
backgrounds, as is stressed in \cite{00436}, 
and we hope to be able to perform it in a separate paper.

\section{Coupled modes}

Unlike $f_{2}$, the functions $f_{0},f_{1}$ and $f_{3}$ 
obey instead a coupled set, 
given by Eqs. (54), (55), (57) of \cite{00436}, which are here written,
more conveniently, in matrix form as (our $L \equiv l(l+1)$, and we 
set $\epsilon=1$ in the Eqs. of \cite{00436}, which corresponds to
studying the vector wave equation (3)) 
\begin{equation}
\left(
\begin{array}{ccc}
P_{00} \quad & A_{3}  \quad& 0 \\
f^{-2}A_{3}  \quad & P_{11}   \quad & 
r^{-2}f^{-1}L C_{3} \\
0   \quad & C_{3}  \quad & P_{33}
\end{array}
\right)
\,
\left(
\begin{array}{l}
f_{0} \\
f_{1} \\
f_{3}
\end{array}
\right)
=0,
\label{(2.1)}
\end{equation}
having defined
\begin{equation}
A_{3} \equiv {2i \Omega H^{3}r \over f},
\label{(2.2)}
\end{equation}
\begin{equation}
C_{3} \equiv {2\over r},
\label{(2.3)}
\end{equation}
\begin{equation}
P_{00} \equiv {d^{2}\over dr^{2}}+Q_{1}{d\over dr}+Q_{2},
\label{(2.4)}
\end{equation}
\begin{equation}
P_{11} \equiv {d^{2}\over dr^{2}}+Q_{3}{d\over dr}+Q_{4},
\label{(2.5)}
\end{equation}
\begin{equation}
P_{33} \equiv {d^{2}\over dr^{2}}+Q_{5}{d\over dr}+Q_{6},
\label{(2.6)}
\end{equation}
\begin{equation}
Q_{1} \equiv C_{3}={2\over r},
\label{(2.7)}
\end{equation}
\begin{equation}
Q_{2} \equiv {\Omega^{2}H^{2}\over f^{2}}-{L \over r^{2}f},
\label{(2.8)}
\end{equation}
\begin{equation}
Q_{3} \equiv {6\over r}\left(1-{2\over 3}{1\over f}\right),
\label{(2.9)}
\end{equation}
\begin{equation}
Q_{4} \equiv {\Omega^{2}H^{2}\over f^{2}}
-\left(4H^{2}+{(L+2) \over r^{2}}\right){1\over f},
\label{(2.10)}
\end{equation}
\begin{equation}
Q_{5} \equiv {2\over r}\left(1-{1\over f}\right),
\label{(2.11)}
\end{equation}
\begin{equation}
Q_{6} \equiv {\Omega^{2}H^{2}\over f^{2}}-{L\over r^{2}}{1\over f},
\label{(2.12)}
\end{equation}
where we have corrected misprints on the right-hand side of Eqs. (54) 
and (55) of \cite{00436}. With our notation, the 
three equations resulting from (8) can be written as
\begin{eqnarray}
\label{(2.13)}
P_{00}f_{0}&=&-A_{3}f_{1},\\
\label{(2.14)}
P_{11}f_{1}&=&-{A_{3}\over f^{2}}f_{0}-{L C_{3}\over r^{2}f}f_{3},\\
\label{(2.15)}
P_{33}f_{3}&=&-C_{3}f_{1}.
\end{eqnarray}

\section{Decoupled equations}

We now express $f_{1}$ from Eq. (20) and we insert it into Eq. 
(21), i.e.
\begin{equation}
P_{11} \left(-{1\over A_{3}}P_{00}f_{0}\right)
=-{A_{3}\over f^{2}}f_{0}-{LC_{3}\over r^{2}f}f_{3}.
\label{(3.1)}
\end{equation}
Next, we exploit the Lorenz gauge condition (2), i.e. \cite{00436}
\begin{equation}
L f_{3}=r^{2}f {d\over dr}\left(-{1\over A_{3}}P_{00}f_{0}\right)
-2r(1-2f)\left(-{1\over A_{3}}P_{00}f_{0}\right)
+i{\Omega H r^{2}\over f}f_{0},
\label{(3.2)}
\end{equation}
and from Eqs. (23) and (24) we obtain, on defining the new independent 
variable $x=rH$, the following fourth-order equation for $f_{0}$:
\begin{equation}
\left[{d^{4}\over dx^{4}}+B_{3}(x){d^{3}\over dx^{3}}
+B_{2}(x){d^{2}\over dx^{2}}+B_{1}(x){d\over dx}
+B_{0}(x) \right]f_{0}(x)=0,
\label{(3.3)}
\end{equation}
where
\begin{equation}
B_{0}(x) \equiv {b_{0}(x)\over x^{4}(x^{2}-1)^{4}},
\label{(3.4)}
\end{equation}
\begin{equation}
b_{0}(x) \equiv L(L-2)+2L(2-L-\Omega^{2})x^{2}
+\Bigr[\Omega^{4}+4 \Omega^{2}+L (L+2(\Omega^{2}-1))\Bigr]x^{4},
\label{(3.5)}
\end{equation}
\begin{equation}
B_{1}(x) \equiv {4(\Omega^{2}+L-2+6x^{2})\over 
x(x^{2}-1)^{2}},
\label{(3.6)}
\end{equation}
\begin{equation}
B_{2}(x) \equiv {2 \Bigr[-L+(\Omega^{2}+L-14)x^{2}+18 x^{4}\Bigr]
\over x^{2}(x^{2}-1)^{2}},
\label{(3.7)}
\end{equation}
\begin{equation}
B_{3}(x) \equiv {4(-1+3x^{2})\over x(x^{2}-1)}.
\label{(3.8)}
\end{equation}
Eventually, $f_{1}$ and $F_{3} \equiv H f_{3}$ 
can be obtained from Eqs. (20) and (24), i.e.
\begin{equation}
f_{1}(x)={i\over 2 \Omega}{(1-x^{2})\over x}
\left({d^{2}\over dx^{2}}
+{2\over x}{d\over dx}+{\Omega^{2}\over (1-x^{2})^{2}}
-{L \over x^{2}(1-x^{2})}\right)f_{0}(x),
\label{(3.9)}
\end{equation}
\begin{equation}
L F_{3}(x)=\left[x^{2}(1-x^{2}) {d\over dx}
-2x(2x^{2}-1)\right]f_{1}(x)
+i\Omega {x^{2}\over (1-x^{2})} f_{0}(x).
\label{(3.10)}
\end{equation}
Our $f_{1}$ and $f_{3}$ are purely imaginary, which means we are
eventually going to take their imaginary part only.
Moreover, as a consistency check, Eqs. (31) and (32) 
have been found to agree with Eq. (22), i.e. (22) is then
identically satisfied.

\section{Exact solutions}

Equation (25) has four linearly independent integrals, so that its
general solution involves four coefficients of linear combination
$C_{1},C_{2},C_{3},C_{4}$, according to (hereafter, $F$ is the
hypergeometric function already used in (6))
\begin{eqnarray}
f_{0}(x)&=& C_{1}x^{-1-l}(1-x^{2})^{-{i\over 2}\Omega}
F \left(-{i \over 2}\Omega-{l\over 2},
-{i \over 2}\Omega+{1\over 2}-{l\over 2};{1\over 2}-l;x^{2}\right)
\nonumber \\
&+& C_{2}x^{-1-l}(1-x^{2})^{-{i\over 2}\Omega}
F \left(-{i \over 2}\Omega+1-{l \over 2},
-{i \over 2}\Omega-{1\over 2}-{l\over 2};
{1\over 2}-l;x^{2} \right) 
\nonumber \\
&+& C_{3}x^{l}(1-x^{2})^{-{i\over 2}\Omega}
F \left(-{i \over 2}\Omega+{l \over 2},
-{i \over 2}\Omega+{3\over 2}+{l\over 2};
{3\over 2}+l;x^{2} \right) 
\nonumber \\
&+& C_{4}x^{l}(1-x^{2})^{-{i\over 2}\Omega}
F \left(-{i \over 2}\Omega+1+{l \over 2},
-{i \over 2}\Omega+{1\over 2}+{l\over 2};
{3\over 2}+l;x^{2} \right).
\label{(4.1)}
\end{eqnarray}
Regularity at the origin ($x=0$ should be included, and we recall
that the event horizon for an observer situated at $x=0$ is given
by $x=1$ \cite{BOUCHER}) 
implies that $C_{1}=C_{2}=0$, and hence, on defining 
\begin{equation}
a_{1} \equiv -{i\over 2}\Omega+{l\over 2}, \;
b_{1} \equiv -{i\over 2}\Omega+{3\over 2}+{l \over 2}, \;
d_{1} \equiv {3\over 2}+l,
\label{(4.3)}
\end{equation}
we now re-express the regular solution in the form (the points
$x=0,1$ being regular singular points of the equation 
(25) satisfied by $f_{0}$)
\begin{equation}
f_{0}(x)=x^{l}(1-x^{2})^{-{i\over 2}\Omega}\Bigr[
C_{3}F(a_{1},b_{1};d_{1};x^{2})
+C_{4}F(a_{1}+1,b_{1}-1;d_{1};x^{2})\Bigr],
\label{(4.4)}
\end{equation}
where the second term on the right-hand side of (35) can be obtained
from the first through the replacements
$$
C_{3} \rightarrow C_{4}, \; 
a_{1} \rightarrow a_{1}+1, \;
b_{1} \rightarrow b_{1}-1,
$$
and the series expressing the two hypergeometric functions are
conditionally convergent, because they satisfy
${\it Re}(c-a-b)=i \Omega$, with
$$
a=a_{1},a_{1}+1; \;
b=b_{1},b_{1}-1; \;
c=d_{1}.
$$
Last, we exploit the identity 
\begin{equation}
{d\over dz}F(a,b;c;z)={ab \over c}F(a+1,b+1;c+1;z)
\label{(4.5)}
\end{equation}
to find, in the formula (31) for $f_{1}(x)$,
\begin{eqnarray}
\; & \; & {d\over dx}f_{0}(x)=C_{3} \biggr \{ 
x^{l-1}(1-x^{2})^{-{i\over 2}\Omega -1}\Bigr[l(1-x^{2})
+i \Omega x^{2}\Bigr]F(a_{1},b_{1};d_{1};x^{2}) \nonumber \\
&+& {2a_{1}b_{1}\over d_{1}}x^{l+1}(1-x^{2})^{-{i\over 2}\Omega}
F(a_{1}+1,b_{1}+1;d_{1}+1;x^{2}) \biggr \} \nonumber \\
&+& \Bigr \{ C_{3} \rightarrow C_{4}, \;
a_{1} \rightarrow a_{1}+1, \;
b_{1} \rightarrow b_{1}-1 \Bigr \}.
\label{(4.6)}
\end{eqnarray}
It is then straightforward, although tedious, to obtain the second 
derivative of $f_{0}$ (see Eq. (38) of Appendix A)
in the equation for $f_{1}$, and the third 
derivative of $f_{0}$ in the formula (32) for $H f_{3}$. The results
are exploited to plot the solutions in Sec. V.

In general, for given initial conditions at $\alpha \in [0,1[$, one can
evaluate $C_{3}$ and $C_{4}$ from
$$
f_{0}(\alpha)=\beta, \;
f_{0}'(\alpha)=\gamma,
$$
i.e. $C_{3}=C_{3}(\beta,\gamma),C_{4}=C_{4}(\beta,\gamma)$.

\section{Plot of the solutions}

To plot the solutions, we begin with $f_{0}$ as given by (35),
which is real-valued despite the many $i$ factors occurring therein.
Figures 1 to 3 describe the solutions for the two choices 
$C_{3}=0,C_{4}=1$ or the other way around, and various values of
$l$ and $\Omega$.

\begin{figure}
\label{fig1}
\includegraphics[scale=0.35]{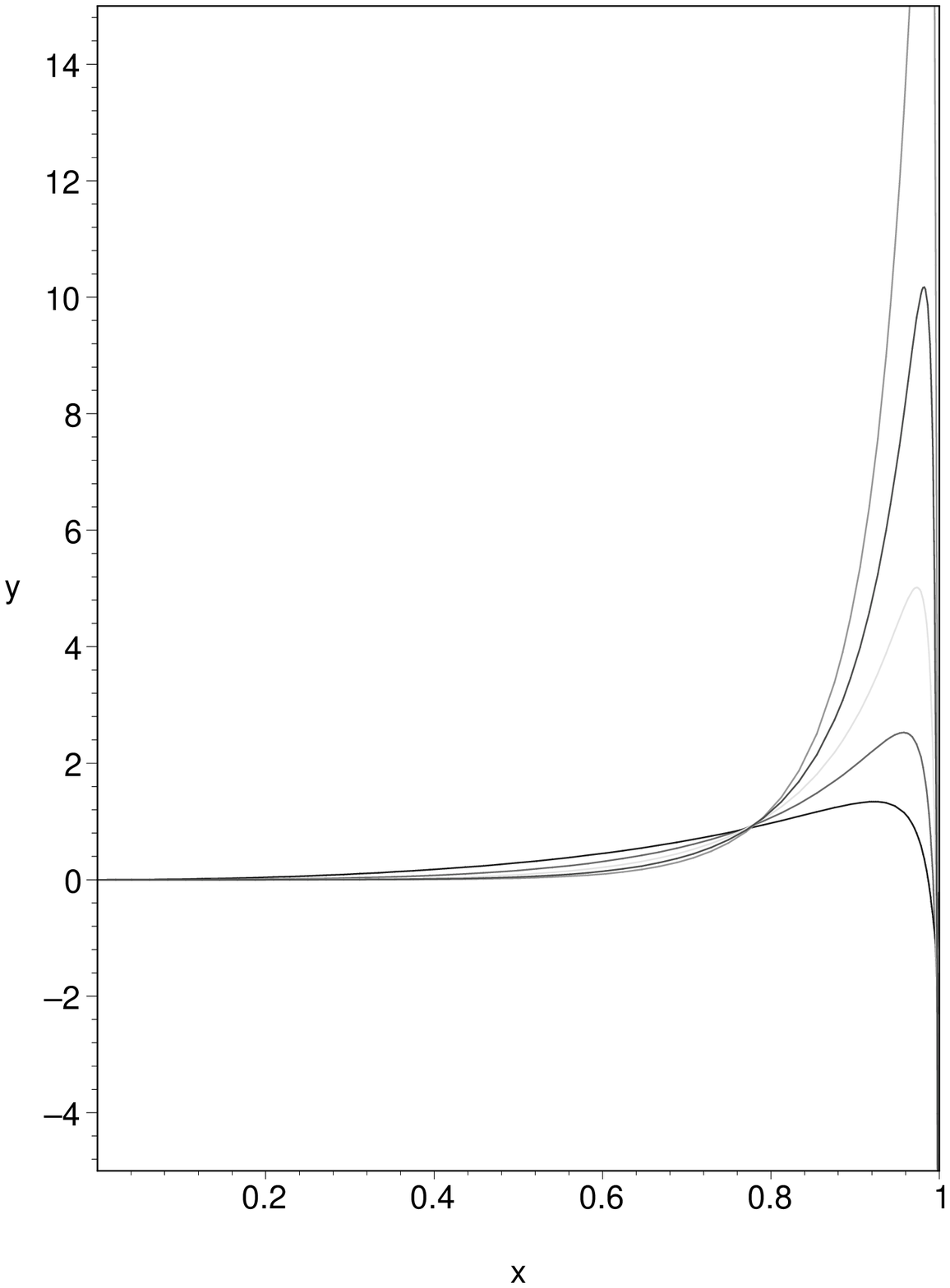}
\includegraphics[scale=0.35]{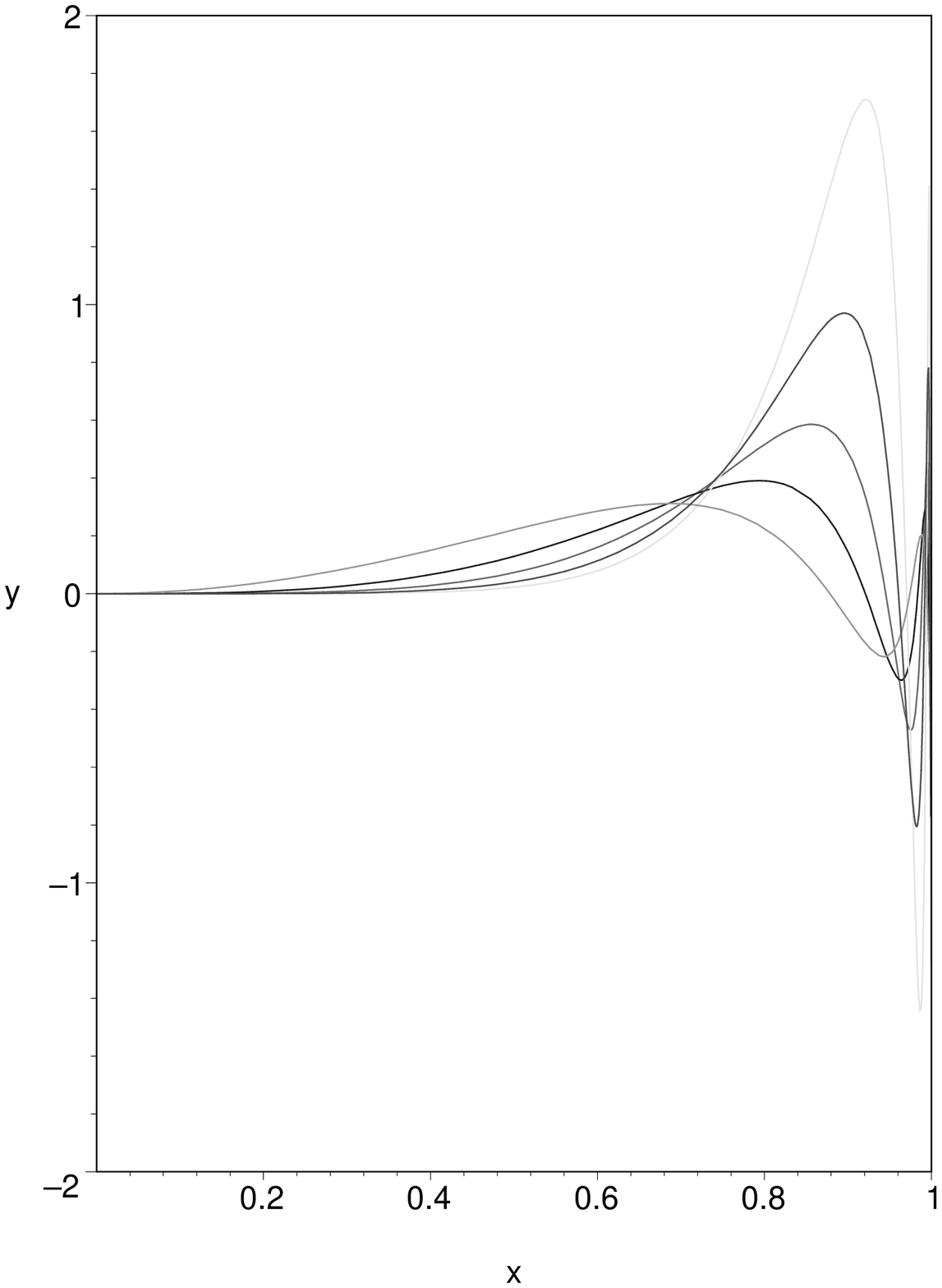}
\caption{Regular solution (35) for $f_{0}$ with $C_{3}=0,C_{4}=1,
l=2,\ldots 6$ with  $\Omega=2$ (left figure) and $\Omega=4$ (right figure). 
Increasing values of $l$ correspond to more peaked curves on the 
right part of the plots.}
\end{figure}

\begin{figure}
\label{fig2}
\includegraphics[scale=0.35]{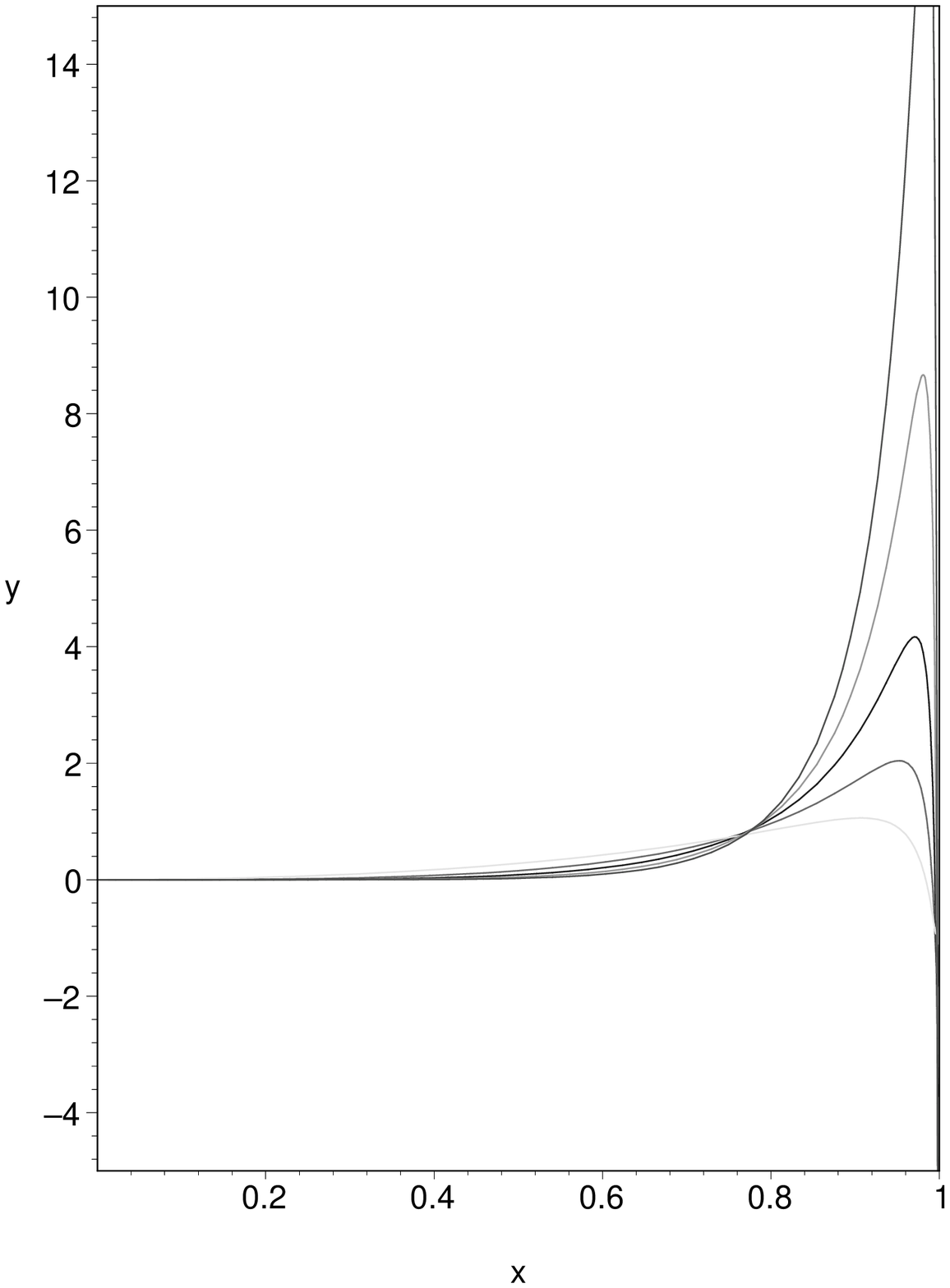}
\includegraphics[scale=0.35]{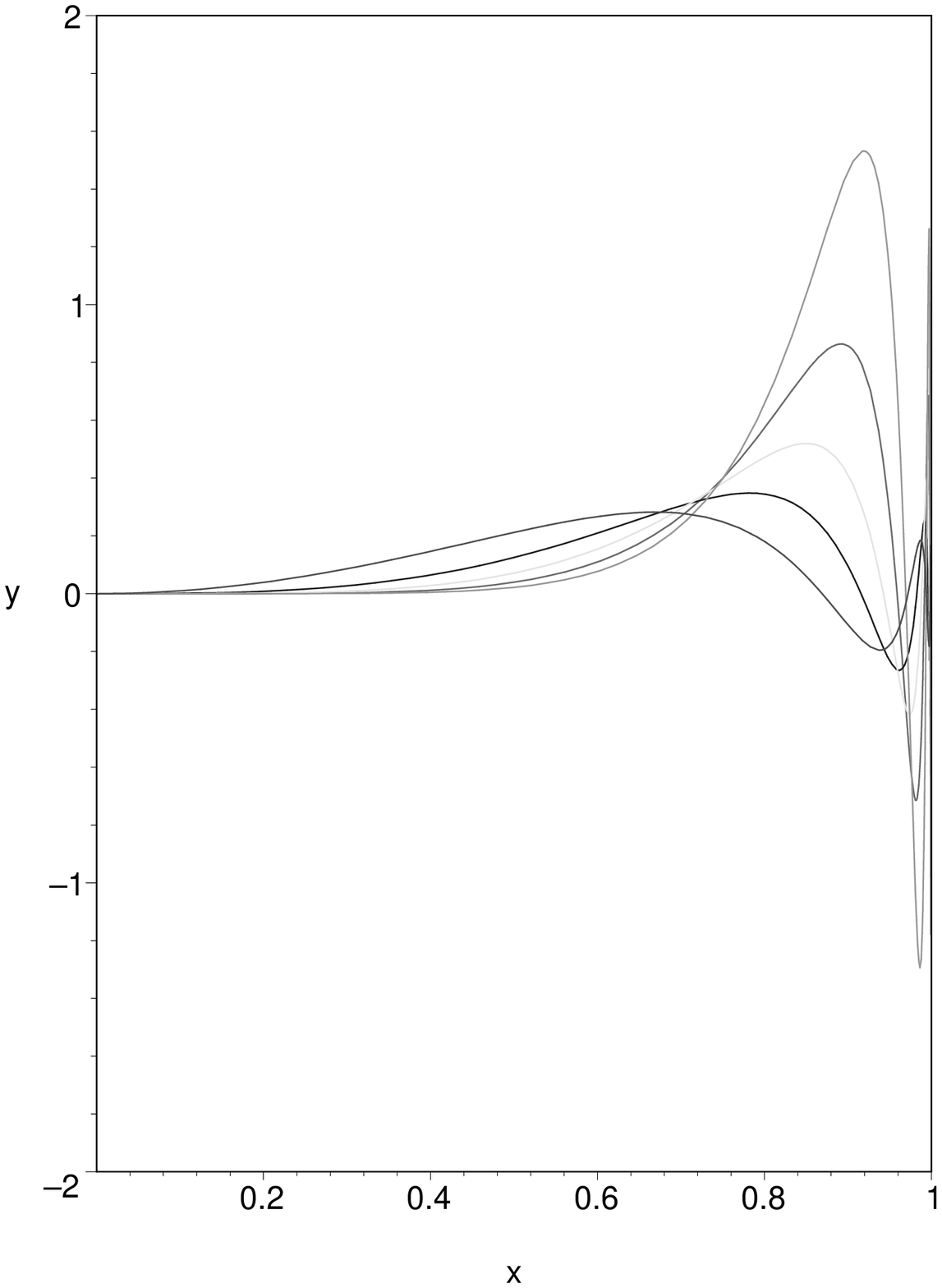}
\caption{Regular solution (35) for $f_{0}$ with $C_{3}=1,C_{4}=0,
l=2,\ldots 6$ with  $\Omega=2$ (left figure) and $\Omega=4$ (right figure).
Increasing values of $l$ correspond to more peaked curves on the 
right part of the plots.}
\end{figure}

\begin{figure}
\label{fig3}
\begin{center}
\includegraphics[scale=0.35]{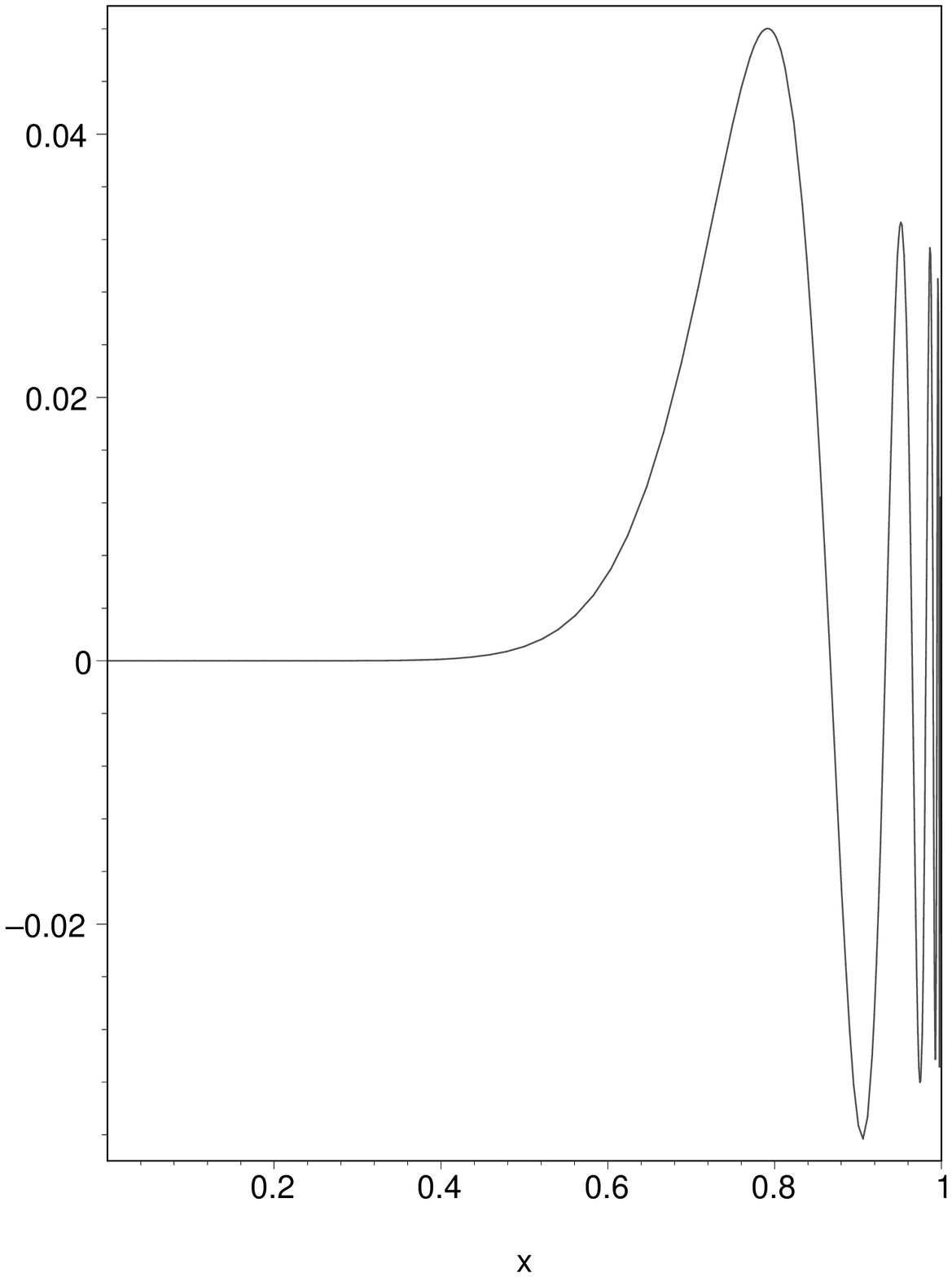}
\end{center}
\caption{Regular solution (35) for $f_{0}$ with $C_{3}=0,C_{4}=1,
l=10$ and $\Omega=10$.}
\end{figure}

We next plot $f_{1}/i$ and $F_{3}/i \equiv H f_{3}/i$ by relying upon
(31) and (32). As far as we can see, all solutions blow up at the
event horizon, corresponding to $x=1$, since there are no static solutions
of the wave equation which are regular inside and on the event horizon
other than the constant one \cite{BOUCHER}.

\begin{figure}
\label{fig4}
\includegraphics[scale=0.35]{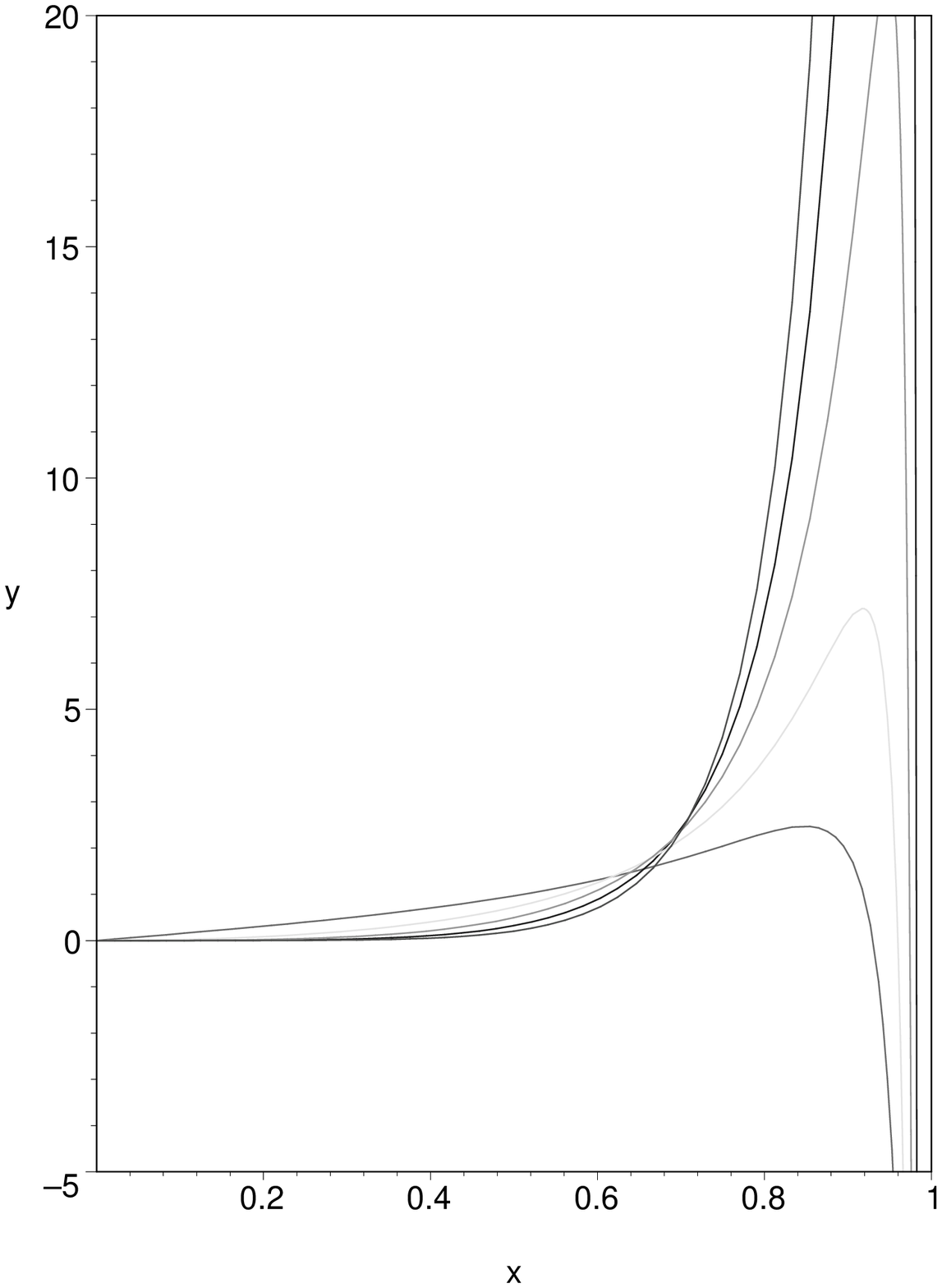}
\includegraphics[scale=0.35]{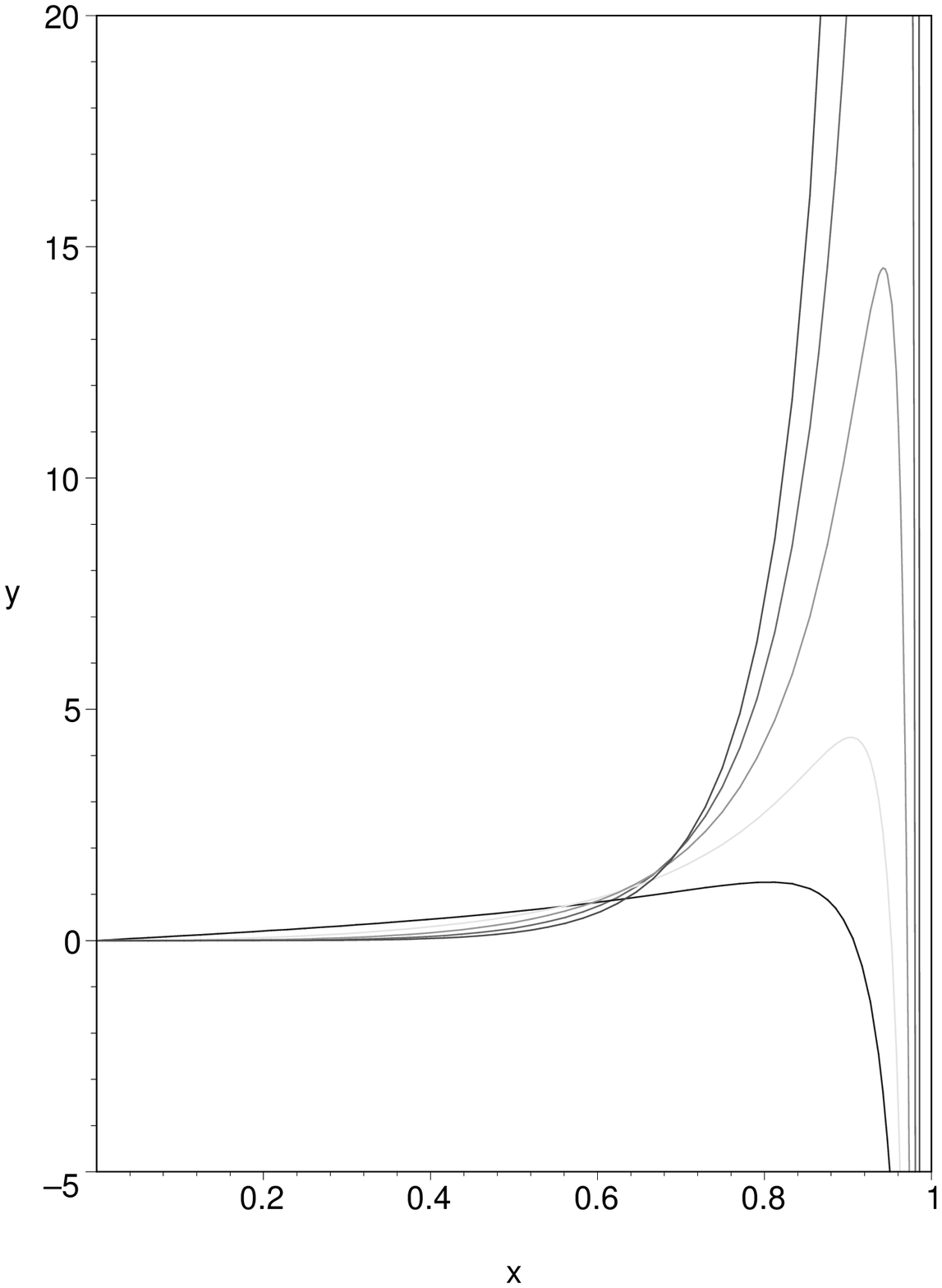}
\caption{Regular solution (31) for $f_{1}/i$ with 
$C_{3}=1,C_{4}=0$ (left figure) and $C_{3}=0,C_{4}=1$ 
(right figure) for $l=2\ldots 6$ and $\Omega=2$.}
\end{figure}

\begin{figure}
\label{fig5}
\includegraphics[scale=0.35]{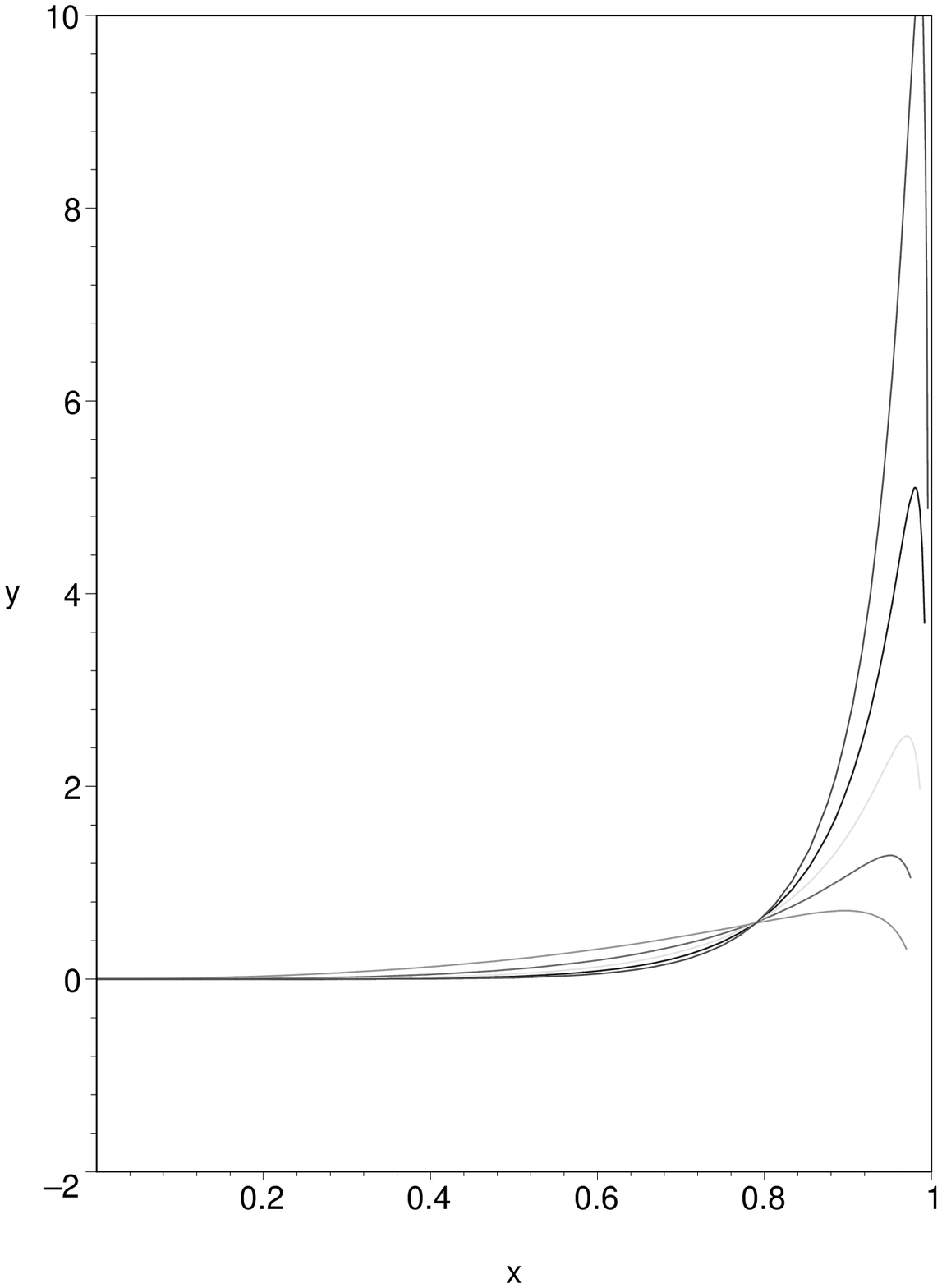}
\includegraphics[scale=0.35]{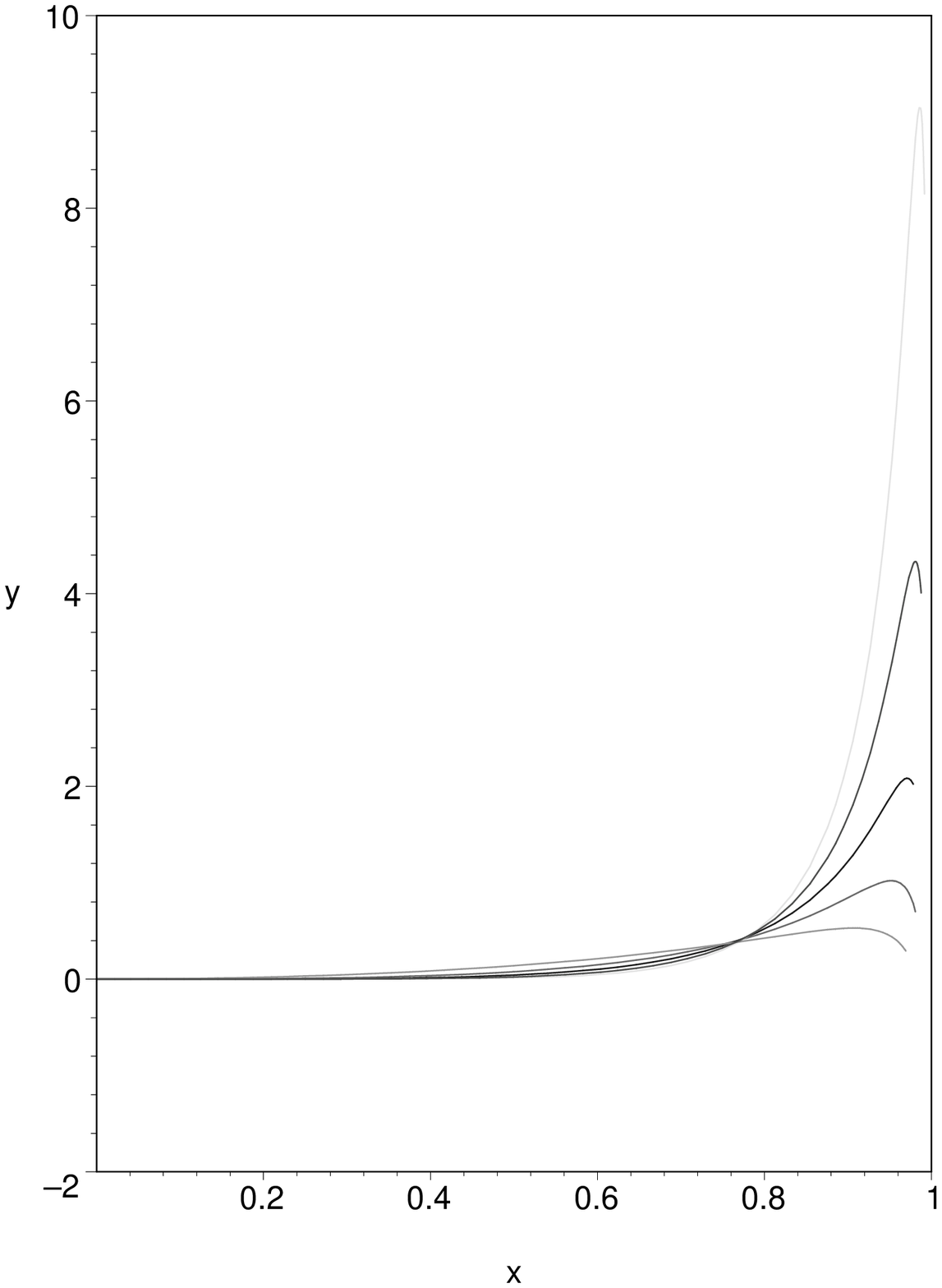}
\caption{Regular solution (32) for $F_{3}/i$ with 
$C_{3}=0,C_{4}=1$ (left figure) and $C_{3}=1,C_{4}=0$ 
(right figure) for $l=2\ldots 6$ and $\Omega=2$.}
\end{figure}

\appendix

\section{Derivatives of $f_{0}$}

The higher-order derivatives of $f_{0}$ in sections III and IV get
increasingly cumbersome, but for completeness we write hereafter 
the result for $f_{0}''(x)$, i.e.
\begin{eqnarray}
{d^{2}\over dx^{2}}f_{0}(x)&=& C_{3} \biggr \{ 
{4a_{1}(a_{1}+1)b_{1}(b_{1}+1)\over d_{1}(d_{1}+1)}x^{l+1}
(1-x^{2})^{-{i\over 2}\Omega} 
F(a_{1}+2,b_{1}+2;d_{1}+2;x^{2}) \nonumber \\
&+& x^{l-1}(1-x^{2})^{-{i\over 2}\Omega-1}
{2a_{1}b_{1}\over d_{1}}x \nonumber \\
& \times & [(2l+1)(1-x^{2})+2i \Omega x^{2}]
F(a_{1}+1,b_{1}+1;d_{1}+1;x^{2}) \nonumber \\
&+& x^{l-2}(1-x^{2})^{-{i\over 2}\Omega-2}
\biggr[l(l-1)(x^{2}-1)^{2}-{i \Omega \over 2}(x^{2}-1)
(lx+2(l+1)x^{2}) \nonumber \\
&+& (2i \Omega- \Omega^{2})x^{4}\biggr]
F(a_{1},b_{1};d_{1};x^{2}) \biggr \} \nonumber \\
&+& \biggr \{ C_{3} \rightarrow C_{4}, \; a_{1} \rightarrow a_{1}+1,
\; b_{1} \rightarrow b_{1}-1 \biggr \}.
\label{(A1)}
\end{eqnarray} 

\section{Special cases: $l=0$ and $l=1$}

We list here the main equations for  $l=0,1$ for completeness. These equations do not add too much to the above discussion and hence we have decided to include them in the appendix.
  
\subsection{The case $l=0, m=0$}

In this case we have $Y(\theta)=(4\pi)^{-1/2}=$ constant 
and the only surviving functions are $f_0$ and $f_1$.
The main equations in \cite{00436} reduce then to (recalling that 
our Eq. (3) requires setting $\epsilon=1$ in \cite{00436})
\beq
\frac{d^2f_0}{dx^2}=-\frac{2}{x}\frac{df_0}{dx}
-\frac{\Omega^{2}}{(x^2-1)^2}f_0
+\frac{2i\Omega x}{(x^2-1)}f_1
\eeq
and the Lorenz gauge condition (24) which now becomes
\beq
\Omega f_0=i(x^2-1)^2\frac{df_1}{dx}+\frac{2i(x^2-1)(2x^2-1)}{x}f_1 \,.
\eeq
These equations can be easily separated and explicitly solved 
in terms of hypergeometric functions. The details are not very illuminating 
and are therefore omitted.

\subsection{The case $l=1, m=0,1$}

In this case we have 
\begin{eqnarray}
Y&=& \sqrt{\frac{3}{4\pi}}\cos \theta \,\quad l=1, m=0\nonumber \\
Y&=& -\sqrt{\frac{3}{8\pi}}\sin \theta \,\quad l=1, m=1\ .
\end{eqnarray}
However, by virtue of the spherical symmetry of the background, 
the equations for both cases $l=1, m=0$ and $l=1, m=1$ do coincide. We have 
\begin{eqnarray}
\frac{d^2f_0}{dx^2}&=&-\frac{2}{x}\frac{df_0}{dx}
-\frac{(\Omega^{2}x^{2}+2x^{2}-2)}
{x^2(x^2-1)^2}f_0+\frac{2i\Omega x}{(x^2-1)}f_{1},\nonumber \\
\frac{d^2f_1}{dx^2}&=&-\frac{2(3x^2-1)}{x(x^2-1)}
\frac{df_1}{dx}-\frac{(\Omega^{2}x^{2}+4x^{4}-4)}
{x^{2}(x^{2}-1)^{2}}f_{1}
+\frac{2i\Omega x}{(x^2-1)^3}f_{0}\nonumber \\
&& +\frac{4F_3}{x^3(x^2-1)}, \nonumber \\
\frac{d^2f_2}{dx^2}&=& -\frac{2(2x^2-1)}{x(x^2-1)}
\frac{df_2}{dx}-\frac{(\Omega^{2}x^{2} +2x^{4} -2)}
{x^2(x^2-1)^2}f_{2},\nonumber \\
\frac{d^2F_3}{dx^2}&=& -\frac{2x}{(x^2-1)}\frac{dF_3}{dx}
-\frac{(\Omega^{2}x^{2}+2x^{2}-2)}
{x^{2}(x^{2}-1)^2}F_{3}-\frac{2}{x}f_{1}.
\end{eqnarray}
To this set one has to add the Lorenz gauge condition 
(24), which now reads
\beq
\Omega f_0=i(x^2-1)^2\frac{df_1}{dx}
+\frac{2i(x^2-1)}{x^2}[F_3+f_1x(2x^2-1)]\,.
\eeq
Once more, the detailed discussion of this case 
can be performed by repeating 
exactly the same steps as in the general case, and is hence omitted.

\section*{Acknowledgments}

G. Esposito is grateful to the Dipartimento di Scienze Fisiche of
Federico II University, Naples, for hospitality and support. 
R. V. Montaquila thanks CNR for partial support. 
D. Bini thanks ICRANet for support.

\end{document}